\documentclass[showpacs,amsmath,amssymb,twocolumn]{revtex4}

\usepackage{graphicx}
\usepackage{dcolumn}
\usepackage{bm}

\begin{document}

\title{\bf\large{Fermion Cooper Pairing with Unequal Masses: Standard Field Theory Approach}}
\author{\normalsize{Lianyi He, Meng Jin and Pengfei Zhuang}}
\affiliation{Physics Department, Tsinghua University, Beijing
100084, China}

\begin{abstract}
The fermion Cooper pairing with unequal masses is investigated in
a standard field theory approach. We derived the superfluid
density and Meissner mass squared of the $U(1)$ gauge field in a
general two species model and found that the often used
proportional relation between the two quantities is broken down
when the fermion masses are unequal. In weak coupling region, the
superfluid density is always negative but the Meissner mass
squared becomes mostly positive when the mass ratio between the
pairing fermions is large enough. We established a proper momentum
configuration of the LOFF pairing with unequal masses and showed
that the LOFF state is energetically favored due to the negative
superfluid density. The single plane wave LOFF state is physically
equivalent to an anisotropic state with a spontaneously generated
superflow. The extension to finite range interaction is briefly
discussed.
\end{abstract}


\maketitle

\section {Introduction}
\label{s1}
The asymmetric cooper pairing between different species of
fermions with mismatched Fermi surfaces, which was discussed many
years ago, promoted new interest in both theoretic and
experimental studies in recent years. The mismatched Fermi
surfaces can be realized, for instance, in a superconductor with
Zeeman splitting induced by an external
field\cite{sarma,larkin,fulde,takada}, an atomic fermion gas
composed of two species of atoms with different densities and/or
masses\cite{exp1,exp2}, an isospin asymmetric nuclear matter with
proton-neutron pairing\cite{sedrakian}, and color superconducting
quark matter with charge neutrality\cite{huang,shovkovy,huang2}.
Among the mechanisms which can produce asymmetry between the
pairing fermions, the mass difference is a very robust one. The
cooper pairing between fermions with unequal masses was firstly
investigated by V.Liu and F.Wilczek\cite{liu}. They considered a
fermion gas composed of light and heavy fermions with attractive
interaction. A homogeneous and isotropic pairing state which is
similar to the Sarma state\cite{sarma} was proposed to be the
ground state of such systems. Such an exotic pairing state is now
called breached pairing (BP) state or interior gap state. In the
BP state there exists gapless fermion excitations, and the
superfluid Fermi gas and the normal Fermi gas coexist in the
momentum space.

It was found many years ago that the Sarma state suffers a
thermodynamic instability\cite{sarma}. It is now generally
accepted that this Sarma instability can be cured in some physical
conditions, such as a long-range interaction when charge
neutrality is required\cite{shovkovy,huang2}, a proper finite
range interaction between the two species of fermions with large
mass difference\cite{forbes}, and a superfluid Fermi gas with
density imbalance in strong coupling
region\cite{pao,son,kitazawa,smit}. While the Sarma instability
can be cured, it was soon found that the superfluid density of the
BP state is negative\cite{wu} and the free energy of the mixed
phase is also lower than that of the BP
state\cite{bedaque,caldas}. Meanwhile, in the study of color
superconductivity, it was found that the gapless color
superconductors possess paramagnetic response to external color
magnetic fields, i.e., the Meissner masses of some gluons are
negative\cite{huang3,huang4,casalbuoni,alford4,fukushima}. All
these phenomena indicate that the homogeneous and isotropic BP
state is unstable and some spatially inhomogeneous and isotropic
states are energetically
favored\cite{giannakis,giannakis2,giannakis3,huang5,hong,gorbar,gorbar2}.
The superfluid density is a fundamental quantity in
superconductivity. It is well known that the superfluid density is
proportional to the Meissner mass squared and often measured via
the London penetration depth in experiments. Due to this relation,
one often regards the negative superfluid density observed in a BP
state and the negative Meissner mass squared observed in a gapless
color superconductor as the same instability.

However, almost all these studies focus on systems where the
masses of the pairing fermions are equal. In the study of
superfluid stability of interior gap states, S.T.Wu and S.Yip
derived a formula of the superfluid density for non-relativistic
asymmetric fermion superfluids with the concept of
quasiparticles\cite{wu}. In the equal mass case, their formula is
consistent with the result calculated from the linear response
theory\cite{fetter}, the current-current correlation
function\cite{nao}, and the Meissner mass squared\cite{he}.
However, in unequal mass systems, the formula of superfluid
density seems quite different from the Meissner mass
squared\cite{he}. Does the proportional relation between the
superfluid density and the Meissner mass squared still hold in
unequal mass systems? In this paper, we will derive the superfluid
density and the Meissner mass squared in unequal mass systems in a
standard field theory approach where the superfluid density and
the Meissner mass squared are treated in the same way.

The paper is organized as follows. In Section \ref{s2}, we briefly
review the formalism of the two species model. In Section
\ref{s3}, we derive the formula of the superfluid density and
compare it with S.T.Wu and S.Yip's phenomenological method. In
Section \ref{s4}, we derive the Meissner mass squared and show
that it is not proportional to the superfluid density in unequal
mass systems. The superfluid density and Meissner mass squared in
the breached pairing states are calculated in Section \ref{s5}.
The LOFF pairing in unequal mass systems is discussed in Section
\ref{s6}. The extension to finite range pairing interaction is
briefly discussed in Section \ref{s7}. We summarize in Section
\ref{s8}. We use the natural unit of $c=\hbar=k_B=1$ through the
paper.

\section { Two Species Model}
\label{s2}
The physical system we are interested in in this paper is an idea
system composed of two species of fermions with attractive
interaction. The system is described by the Lagrangian density
with imaginary time $\tau=it$,
\begin{eqnarray}
\label{lagrangian} {\cal
L}=\sum_{i=a,b}\psi_{i}^*\left(-\partial_\tau+\frac{\nabla^2}{2m_i}+\mu_i\right)\psi_{i}+g\psi_{a}^*\psi_{b}^*\psi_{b}\psi_{a},
\end{eqnarray}
where $\psi_i\equiv\psi_i(x)$ with $x=(\tau,\vec{x})$ are fermion
fields for the two species $a$ and $b$, the coupling constant $g$
is positive to keep the interaction attractive, $m_a$ and $m_b$
are the masses for the two species, and $\mu_a$ and $\mu_b$ the
chemical potentials.

The key quantity to describe a thermodynamic system is the
partition function $Z$. It can be expressed as
\begin{equation}
Z=\int[d\psi_i][d\psi^*_i]e^{\int_0^\beta d\tau\int d^3{\bf
x}{\cal L}[\psi_i,\psi^*_i]}
\end{equation}
in the imaginary time formalism of finite temperature field
theory, where $\beta$ is the inverse of the temperature $T$,
$\beta=1/T$. For attractive interaction $g$, we can perform an
exact Hubbard-Stratonovich transformation to introduce the
auxiliary boson field $\phi(x)$ and its complex conjugate
$\phi^*(x)$. With the Nambu-Gorkov fields $\Psi,\bar{\Psi}$
defined as
\begin{equation}
\Psi(x)=\left(\begin{array}{c} \psi_a
\\ \psi_b^* \end{array}\right)\ ,\ \ \ \bar{\Psi}(x)=\left(\begin{array}{cc}
\psi^*_a& \psi_b\end{array}\right)\ ,
\end{equation}
we can express the partition function as
\begin{equation}
Z=\int[d\Psi][d\bar{\Psi}][d\phi][d\phi^*]e^{\int_0^\beta
d\tau\int d^3{\bf x}\left(\bar{\Psi}{\cal
K}\Psi-|\phi|^2/g\right)},
\end{equation}
where the kernel ${\cal K}[\phi,\phi^*]$ is defined as
\begin{equation}
{\cal K}[\phi,\phi^*]=\left(\begin{array}{cc}
-\partial_\tau+\frac{\nabla^2}{2m_a}+\mu_a&\phi
\\ \phi^*&-\partial_\tau-\frac{\nabla^2}{2m_b}-\mu_b\end{array}\right)\ .
\end{equation}

In mean field approximation, we replace $\phi$ and $\phi^*$ by
their ensemble averages $\Delta$ and $\Delta^*$. In a homogenous
and isotropic state, they are independent of coordinates. Then we
can directly evaluate the Gaussian path integral and obtain the
thermodynamic potential
\begin{equation}
\Omega =\frac{|\Delta|^2}{g}-T\sum_n\int\frac{d^3{\bf
p}}{(2\pi)^3}\textrm{Tr} \ln {\cal G}^{-1}(i\omega_n,{\bf p})
\end{equation}
in terms of the inverse fermion propagator
\begin{equation}
{\cal G}^{-1}(i\omega_n,{\bf p})=\left(\begin{array}{cc}
i\omega_n-\epsilon_{\bf p}^a&\Delta
\\ \Delta^*&i\omega_n+\epsilon_{\bf p}^b\end{array}\right)
\end{equation}
with the free fermion dispersions $\epsilon^i_{\bf p}={\bf
p}^2/(2m_i)-\mu_i$. The explicit form of the fermion propagator
which we need in the following sections can be explicitly
expressed as
\begin{equation}
{\cal G}(i\omega_n,{\bf p})=\left(\begin{array}{cc} {\cal
G}_{11}(i\omega_n,{\bf p}) &{\cal G}_{12}(i\omega_n,{\bf p})
\\ {\cal G}_{21}(i\omega_n,{\bf p})&{\cal G}_{22}(i\omega_n,{\bf p})\end{array}\right)
\end{equation}
with the matrix elements
\begin{eqnarray}
{\cal G}_{11}(i\omega_n,{\bf p}) &=&
{i\omega_n-\epsilon_A+\epsilon_S\over (i\omega_n-\epsilon_A)^2-\epsilon_\Delta^2},\nonumber\\
{\cal G}_{22}(i\omega_n,{\bf p}) &=&
{i\omega_n-\epsilon_A-\epsilon_S\over
(i\omega_n-\epsilon_A)^2-\epsilon_\Delta^2},\nonumber\\
{\cal G}_{12}(i\omega_n,{\bf p}) &=& {-\Delta\over
(i\omega_n-\epsilon_A)^2-\epsilon_\Delta^2},\nonumber\\
{\cal G}_{21}(i\omega_n,{\bf p}) &=& {-\Delta^*\over
(i\omega_n-\epsilon_A)^2-\epsilon_\Delta^2},
\end{eqnarray}
where the quantities $\epsilon_S,\epsilon_A$ and $\epsilon_\Delta$
are defined as
\begin{equation}
\epsilon_{S,A}=(\epsilon_{\bf p}^a\pm\epsilon_{\bf p}^b)/2,\ \ \
\epsilon_\Delta=\sqrt{\epsilon_S^2+|\Delta|^2}.
\end{equation}
Since all quantities depend only on $|\Delta|$, we can set
$\Delta$ to be real from now on. From the pole of the fermion
propagator we can read the dispersions $\epsilon_{\bf p}^A$ and
$\epsilon_{\bf p}^B$ of fermionic quasiparticles:
\begin{eqnarray}
\epsilon_{\bf p}^A=\epsilon_\Delta+\epsilon_A,\ \ \ \epsilon_{\bf
p}^B=\epsilon_\Delta-\epsilon_A\ .
\end{eqnarray}
For $\epsilon_A=0$, we recover the well know BCS type excitation.
The asymmetric part $\epsilon_A$ is the key quantity to produce
exotic pairing states.

The occupation numbers of the two species of fermions can be
calculated via the diagonal elements of the fermion propagator,
\begin{eqnarray}
n_a({\bf p})&=&T\lim_{\eta\to 0}\sum_n{\cal
G}_{11}(i\omega_n,{\bf p})e^{i\omega_n\eta},\nonumber\\
n_b({\bf p})&=&-T\lim_{\eta\to 0}\sum_n{\cal
G}_{22}(i\omega_n,{\bf p})e^{-i\omega_n\eta}.
\end{eqnarray}
Completing the Matsubara frequency summation, we obtain
\begin{eqnarray}
n_a({\bf p})&=&u_p^2f(\epsilon_{\bf p}^A)
+v_p^2f(-\epsilon_{\bf p}^B),\nonumber\\
n_b({\bf p})&=&u_p^2f(\epsilon_{\bf p}^B)+v_p^2f(-\epsilon_{\bf
p}^A)\
\end{eqnarray}
with the coherent coefficients
$u_p^2=\left(1+\epsilon_S/\epsilon_\Delta\right)/2$ and
$v_p^2=\left(1-\epsilon_S/\epsilon_\Delta\right)/2$. The particle
number densities $n_a$ and $n_b$ for the species $a$ and $b$ are
obtained by integrating $n_a({\bf p})$ and $n_b({\bf p})$ over
momentum.

\section {Superfluid Density}
\label{s3}
In this section we try to derive the superfluid density in a
standard field theory approach. When the superfluid moves with a
uniform but small velocity ${\bf v}_s$, the condensates transform
as $\Delta\rightarrow\Delta e^{2i{\bf q}\cdot{\bf x}}$ and
$\Delta^*\rightarrow\Delta^* e^{-2i{\bf q}\cdot{\bf x}}$ with the
total momentum of the cooper pair $2{\bf q}=(m_a+m_b){\bf v}_s$,
and the fermion fields transform as
$\psi_a\rightarrow\psi_ae^{i{\bf q}_a\cdot{\bf x}}$ and
$\psi_b\rightarrow\psi_be^{i{\bf q}_b\cdot{\bf x}}$ with the
momenta of the two species ${\bf q}_a=m_a{\bf v}_s$ and ${\bf
q}_b=m_b{\bf v}_s$ which satisfy ${\bf q}_a+{\bf q}_b=2{\bf q}$.
The superfluid density tensor $\rho_{ij}$ is defined as\cite{lida}
\begin{equation}
\Omega({\bf v}_s)=\Omega({\bf 0})+{\bf j}_s\cdot{\bf
v}_s+\frac{1}{2}\rho_{ij}({\bf v}_s)_i({\bf v}_s)_j+\cdots.
\end{equation}
For a homogeneous and isotropic superfluid, we have
$\rho_{ij}=\delta_{ij}\rho_s/3$, and the above formula can be
reduced to
\begin{equation}
\Omega({\bf v}_s)=\Omega({\bf 0})+{\bf j}_s\cdot{\bf
v}_s+\frac{1}{2}\rho_s{\bf v}_s^2+\cdots,
\end{equation}
where $\rho_s$ is the superfluid density. When $\rho_s$ is
negative, the homogeneous and isotropic state is unstable and a
state with spontaneously generated superflow which breaks the
rotational symmetry is energetically favored.

After the transformation of the condensates and the fermion
fields, the thermodynamic potential is changed as
\begin{equation}
\Omega({\bf v}_s) =\frac{\Delta^2}{g}-T\sum_n\int\frac{d^3{\bf
p}}{(2\pi)^3}\textrm{Tr} \ln {\cal G}_s^{-1}(i\omega_n,{\bf p})
\end{equation}
in terms of the ${\bf v}_s$-dependent inverse propagator
\begin{equation}
{\cal G}_s^{-1}(i\omega_n,{\bf p})=\left(\begin{array}{cc}
i\omega_n-\epsilon_{{\bf p}+{\bf q}_a}^a&\Delta
\\ \Delta&i\omega_n+\epsilon_{{\bf p}-{\bf q}_b}^b\end{array}\right).
\end{equation}
Using the relation
\begin{equation}
{\cal G}_s^{-1}={\cal G}^{-1}-{\bf p}\cdot{\bf v}_s-{1\over
2}\Sigma_m{\bf v}_s^2
\end{equation}
with the matrix $\Sigma_m=diag(m_a,-m_b)$, we can do the
derivative expansion
\begin{eqnarray}
\textrm{Tr}\ln {\cal G}_s^{-1}-\textrm{Tr}\ln {\cal G}^{-1}
&=&{\bf p}\cdot{\bf v}_s \textrm{Tr}\left({\cal G}\right)-{{\bf
v}_s^2\over 2}\textrm{Tr}\left({\cal G}\Sigma_m\right)
\nonumber\\
&-&\frac{1}{2}({\bf p}\cdot{\bf v}_s)^2 \textrm{Tr}\left({\cal
G}{\cal G}\right)+\cdots.
\end{eqnarray}
With this expansion, we can expand $\Omega({\bf v}_s)$ in powers
of ${\bf v}_s$. The superfluid density can be read from the
quadratic term in ${\bf v}_s$. After some direct algebras, we
obtain
\begin{eqnarray}
\rho_s=m_an_a+m_bn_b+\int\frac{d^3{\bf
p}}{(2\pi)^3}\frac{p^2}{3}\left({\cal T}_{11}+{\cal T}_{22}+2{\cal
T}_{12}\right),
\end{eqnarray}
where ${\cal T}_{11}, {\cal T}_{22}, {\cal T}_{12}$ are the
fermion Matsubara frequency summations defined as
\begin{eqnarray}
{\cal T}_{11}&=&T\sum_n{\cal G}_{11}{\cal
G}_{11}\\
&=&u_p^2v_p^2\frac{f(\epsilon_{\bf p}^A)+f(\epsilon_{\bf p}^B)-1}
{\epsilon_\Delta}+u_p^4f^\prime(\epsilon_{\bf p}^A)+v_p^4f^\prime(\epsilon_{\bf p}^B), \nonumber\\
{\cal T}_{22}&=&T\sum_n{\cal G}_{22}{\cal
G}_{22}\nonumber\\
&=&u_p^2v_p^2\frac{f(\epsilon_{\bf p}^A)+f(\epsilon_{\bf p}^B)-1}
{\epsilon_\Delta}+v_p^4f^\prime(\epsilon_{\bf p}^A)+u_p^4f^\prime(\epsilon_{\bf p}^B)\ , \nonumber\\
{\cal T}_{12}&=&T\sum_n{\cal G}_{12}{\cal G}_{21}\nonumber\\
&=&u_p^2v_p^2\left[\frac{1-f(\epsilon_{\bf p}^A)-f(\epsilon_{\bf
p}^B)} {\epsilon_\Delta}+f^\prime(\epsilon_{\bf
p}^A)+f^\prime(\epsilon_{\bf p}^B)\right]\nonumber
\end{eqnarray}
with $f(x)$ being the Fermi distribution function and
$f^\prime(x)=df(x)/dx$. Using these results we get
\begin{eqnarray}
\rho_s=m_an_a+m_bn_b+\int{d^3{\bf p}\over (2\pi)^3}{p^2\over
3}\left[f^\prime(\epsilon_{\bf p}^A)+f^\prime(\epsilon_{\bf
p}^B)\right].
\end{eqnarray}
One can easily check that this formula is invariant under the
exchange $a\leftrightarrow b$. When $\Delta=0$, i.e., in the
normal state, we have
\begin{eqnarray}
\rho_s&=&\int_0^\infty dp{p^2\over
2\pi^2}\left[m_af(\epsilon_{\bf p}^a)+m_bf(\epsilon_{\bf p}^b)\right]\nonumber\\
&+&\int_0^\infty dp{p^4\over 6\pi^2}\left[f'(\epsilon_{\bf
p}^a)+f'(\epsilon_{\bf p}^b)\right].
\end{eqnarray}
From the identity
\begin{equation}
\int_0^\infty dp{p^4\over 6\pi^2}f'\left(\epsilon^i_{\bf
p}\right)=-m_i\int_0^\infty dp{p^2\over
2\pi^2}f\left(\epsilon^i_{\bf p}\right),
\end{equation}
$\rho_s$ vanishes automatically in the normal state.

The result we obtained here is in agreement with the formula
derived by S.T.Wu and S.Yip with a phenomenological
method\cite{wu}. With their method, in presence of a small
superfluid velocity ${\bf v}_s$, the quasiparticle energies are
shifted by ${\bf p}\cdot{\bf v}_s$ and the occupation numbers
become
\begin{eqnarray}
\tilde{n}_a({\bf p})&=&u_p^2f(\epsilon_{\bf p}^A+{\bf p}\cdot{\bf
v}_s)
+v_p^2f(-\epsilon_{\bf p}^B+{\bf p}\cdot{\bf v}_s),\nonumber\\
\tilde{n}_b({\bf p})&=&u_p^2f(\epsilon_{\bf p}^B+{\bf p}\cdot{\bf
v}_s)+v_p^2f(-\epsilon_{\bf p}^A+{\bf p}\cdot{\bf v}_s).
\end{eqnarray}
The number current can be decomposed into a diamagnetic and a
paramagnetic parts:
\begin{eqnarray}
&&{\bf J}^d_i=\int{d^3{\bf p}\over
(2\pi)^3}\tilde{n}_i({\bf p}){\bf v}_s\equiv\rho_i^d{\bf v}_s,\nonumber\\
&&{\bf J}^p_i=\frac{1}{m_i}\int{d^3{\bf p}\over (2\pi)^3}{\bf
p}\tilde{n}_i({\bf p})\equiv\rho_i^p{\bf v}_s.
\end{eqnarray}
To leading order in ${\bf v}_s$, we have $\rho^d_i=n_i$. Using the
fact ${\bf J}_i^p=0$ for ${\bf v}_s=0$, i.e.,
\begin{equation}
{\bf J}^p_i=\frac{1}{m_i}\int{d^3{\bf p}\over (2\pi)^3}{\bf
p}(\tilde{n}_i({\bf p})-n_i({\bf p})),
\end{equation}
we obtain
\begin{eqnarray}
&&\rho_a^p=\frac{1}{m_a}\int{d^3{\bf
p}\over (2\pi)^3}{p^2\over 3}\left[u_p^2f^\prime(\epsilon_{\bf p}^A)+v_p^2f^\prime(\epsilon_{\bf p}^B)\right],\nonumber\\
&&\rho_b^p=\frac{1}{m_b}\int{d^3{\bf p}\over (2\pi)^3}{p^2\over
3}\left[u_p^2f^\prime(\epsilon_{\bf
p}^B)+v_p^2f^\prime(\epsilon_{\bf p}^A)\right].
\end{eqnarray}
The total superfluid density is defined as
$\rho_s=m_a\rho_a+m_b\rho_b$ with $\rho_i=\rho_i^d+\rho_i^p$.
Using the fact $u_p^2+v_p^2=1$, it is exactly the formula we
obtained above. From our derivation, we can also decompose the
superfluid density into two parts
\begin{eqnarray}
&&\rho_s^a=m_an_a+\int\frac{d^3{\bf
p}}{(2\pi)^3}\frac{p^2}{3}\left({\cal T}_{11}+{\cal
T}_{12}\right),\nonumber\\
&&\rho_s^b=m_bn_b+\int\frac{d^3{\bf
p}}{(2\pi)^3}\frac{p^2}{3}\left({\cal T}_{22}+{\cal
T}_{12}\right),
\end{eqnarray}
where $\rho_s^a\equiv m_a\rho_a$ and $\rho_s^b\equiv m_b\rho_b$
can be defined as the superfluid densities for the two species of
fermions.

One may ask why these two derivations give the same result. In
presence of a small superflow ${\bf v}_s$, the quasiparticle
dispersions can be read from $\det{\cal G}_s^{-1}=0$. After a
simple algebra, we get the modified dispersions for the
quasiparticles
\begin{eqnarray}
\tilde{\epsilon}_{\bf p}^A&=&\sqrt{\tilde{\epsilon}_S^2+\Delta^2}+\tilde{\epsilon}_A+{\bf p}\cdot{\bf v}_s,\nonumber\\
\tilde{\epsilon}_{\bf
p}^B&=&\sqrt{\tilde{\epsilon}_S^2+\Delta^2}-\tilde{\epsilon}_A-{\bf
p}\cdot{\bf v}_s
\end{eqnarray}
with
\begin{eqnarray}
\tilde{\epsilon}_S=\epsilon_S+\frac{1}{4}(m_a+m_b){\bf v}_s^2,\nonumber\\
\tilde{\epsilon}_A=\epsilon_A+\frac{1}{4}(m_a-m_b){\bf v}_s^2.
\end{eqnarray}
To leading order in ${\bf v}_s$, the quasiparticle energies are
really shifted by ${\bf p}\cdot{\bf v}_s$. However, one should
note that the derivation with the concept of quasiparticle is
inconsistent. The particle occupation numbers in presence of a
superflow should be
\begin{eqnarray}
\tilde{n}_a({\bf p})&=&u_p^2f(\epsilon_{\bf p}^A+{\bf p}\cdot{\bf
v}_s)
+v_p^2f(-\epsilon_{\bf p}^B+{\bf p}\cdot{\bf v}_s),\nonumber\\
\tilde{n}_b({\bf p})&=&u_p^2f(\epsilon_{\bf p}^B-{\bf p}\cdot{\bf
v}_s)+v_p^2f(-\epsilon_{\bf p}^A-{\bf p}\cdot{\bf v}_s)
\end{eqnarray}
to leading order in ${\bf v}_s$. Using this correct occupation
numbers, one can not obtain the correct result since $\rho^p_b$
will change a sign. We guess that for such an asymmetric system,
one can not self-consistently derive the superfluid density with
the concept of quasiparticles. Only for symmetric systems like the
standard BCS, the method works as discussed in text books.

The formula of the superfluid density we derived here is in
principle only suitable for grand canonical ensembles where the
chemical potentials $\mu_a$ and $\mu_b$ are fixed\cite{liu2}. For
systems where the particle numbers $n_a$ and $n_b$ are fixed or
the total number $n=n_a+n_b$ is fixed, we should use the free
energy ${\cal F}=\Omega+\mu_an_a+\mu_bn_b$ or ${\cal
F}=\Omega+(\mu_a+\mu_b)n/2$ instead of the thermodynamic potential
$\Omega$ to calculate the superfluid density. However, we can show
that such a correction is beyond order $O({\bf v}_s^2)$ for an
homogeneous and isotropic state\cite{wu2}, and hence we can safely
apply the above formula to the systems with fixed particle
numbers. For instance, if $n_a$ and $n_b$ is fixed, we have
\begin{eqnarray}
\rho_s=\frac{\partial^2{\cal F}}{\partial {\bf v}_s^2}\bigg|_{{\bf
v}_s=0}=\frac{\partial^2\Omega}{\partial {\bf v}_s^2}\bigg|_{{\bf
v}_s=0}+\frac{\partial n_i}{\partial {\bf v}_s}\frac{\partial
\mu_i}{\partial {\bf v}_s}\bigg|_{{\bf v}_s=0}.
\end{eqnarray}
For a homogeneous and isotropic state, the term $\partial
n_i/\partial {\bf v}_s|_{{\bf v}_s=0}$ vanishes automatically.

In our derivation, we did not use the assumption of weak coupling,
and the formula can be applied to study the superfluid stability
in the BCS-BEC crossover in a light-heavy fermion gas such as a
mixture of $^6$Li and $^{40}$K. In recent studies on BCS-BEC
crossover in equal mass systems, it was found that the BP state is
stable in the BEC region, i.e., it is free from the Sarma
instability and negative superfluid density\cite{pao}. We expect
that such a stable BP state can also be realized in a light-heavy
fermion gas in strong coupling.

\section {Meissner Mass}
\label{s4}
The two species model is invariant under the following phase
transformation
\begin{eqnarray}
\psi_{i}(x)\rightarrow e^{i\varphi_i}\psi_{i}(x), \ \ \
\phi(x)\rightarrow e^{i(\varphi_a+\varphi_b)}\phi(x)
\end{eqnarray}
with arbitrary and constant phases $\varphi_a$ and $\varphi_b$,
which means that the symmetry group of the model is
$U(1)_{\varphi_a}\otimes U(1)_{\varphi_b}$. The order parameter is
invariant only for $\varphi_a=-\varphi_b$. In presence of a
nonzero expectation value of $\phi$, the symmetry group is
spontaneously broken down to a $U(1)$ subgroup
\begin{eqnarray}
U(1)_{\varphi_a}\otimes U(1)_{\varphi_b}\rightarrow
U(1)_{\varphi_a-\varphi_b}.
\end{eqnarray}
The unbroken $U(1)$ subgroup corresponds to the phase difference
$\Delta\varphi=\varphi_a-\varphi_b$, and a Goldstone mode
corresponding to the total phase $\varphi=\varphi_a+\varphi_b$
will appear. Let's add a $U(1)$ gauge field $A_\mu$ in the
Lagrangian,
\begin{equation}
\label{lag} {\cal L}=\sum_{i=a,b}\psi_{i}^*\left(-D_{\tau
i}+\frac{{\bf
D}_i^2}{2m_i}+\mu_i\right)\psi_{i}+g\psi_{a}^*\psi_{b}^*\psi_{b}\psi_{a}+{\cal
L}_A
\end{equation}
with $D_{\mu i}=\partial_\mu-ieQ_i A_\mu$, where ${\cal L}_A$ is
the gauge field sector, and $eQ_a,eQ_b$ are the gauge couplings
for the two species of fermions. In presence of a gauge field, the
Goldstone mode disappears and the gauge field will obtain a mass
$m_A$ via Higgs mechanism. This is called Meissner effect in
superconductivity, and the mass $m_A$ the gauge field obtains is
called Meissner mass.

The standard way to calculate the Meissner mass is to evaluate the
polarization tensor $\Pi_{\mu\nu}(K)$ of the gauge field. For the
interaction (\ref{lag}), the spatial components of the
polarization tensor read
\begin{equation}
\Pi_{ij}(k)=\Pi^d_{ij}(k)+\Pi^p_{ij}(k)
\end{equation}
with the diamagnetic part
\begin{equation}
\Pi^d_{ij}(k)=\delta_{ij}e^2\frac{T}{V}\sum_p\textrm{Tr}[{\cal
G}(p)\Sigma_d]
\end{equation}
and the paramagnetic part
\begin{eqnarray}
\Pi^p_{ij}(k)=e^2\frac{T}{V}\sum_p p_ip_j\textrm{Tr}\left[{\cal
G}\left(p_+\right)\Sigma_p{\cal G}\left(p_-\right)\Sigma_p\right]
\end{eqnarray}
with $p_\pm=p\pm k/2$, where the matrices $\Sigma_d$ and
$\Sigma_p$ are defined as
\begin{equation}
\Sigma_d=\left(\begin{array}{cc} \frac{Q_a^2}{m_a}&0
\\ 0&-\frac{Q_b^2}{m_b}\end{array}\right),\
 \Sigma_p=\left(\begin{array}{cc} \frac{Q_a}{m_a}&0
\\ 0&\frac{Q_b}{m_b}\end{array}\right).
\end{equation}
Note that the paramagnetic part is just the current-current
correlation function which gives both diamagnetic part and
paramagnetic part in relativistic systems\cite{huang3,huang4} but
only paramagnetic part in non-relativistic systems. The Meissner
mass $m_A$ can be evaluated by
\begin{equation}
m_A^2=\frac{1}{2}\lim_{{\bf
k}\rightarrow0}(\delta_{ij}-\hat{k}_i\hat{k}_j)\Pi_{ij}(0,{\bf
k}).
\end{equation}
If $m_A^2$ is negative, the homogeneous and isotropic state
suffers the magnetic instability\cite{huang3,huang4} and a state
with gauge field condensation $\langle{\bf A}\rangle\neq 0$ which
breaks the rotational symmetry is energetically favored.

For a clear comparison of the superfluid density and the Meissner
mass squared in unequal mass systems, we employ another
approach\cite{he}. The Meissner mass squared can be calculated via
the response of the effective potential to an external transverse
vector potential. In presence of a small external vector potential
${\bf A}(0,{\bf q}\rightarrow 0)$ in the static and long wave
limit, the effective potential of the system can be expanded in
powers of ${\bf A}$,
\begin{equation}
\Omega({\bf A})=\Omega({\bf 0})+{\bf J}_A\cdot{\bf A}+\frac{1}{2}
M_{ij}^2{\bf A}\cdot {\bf A}+\ldots
\end{equation}
with the coefficients
\begin{equation}
M^2_{ij}=\frac{\partial^2 \Omega({\bf A})}{
\partial A_i\partial A_j}\Big|_{\bf A=0}\ .
\end{equation}
The coefficients $M^2_{ij}$ are just the components of the
Meissner mass squared tensor. In a homogenous and isotropic
superconductor, we have $M^2_{ij}=0$ for $i\neq j$ and
$M^2_{11}=M^2_{22}=M^2_{33}$, and the Meissner mass squared
$m_A^2$ is defined as
\begin{equation}
m_A^2=\frac{1}{3}\sum_{i=1}^3M_{ii}^2.
\end{equation}
The thermodynamic potential in presence of the static and long
wave vector potential ${\bf A}$ can be expressed as
\begin{equation}
\Omega({\bf A}) =\frac{\Delta^2}{g}-T\sum_n\int\frac{d^3{\bf
p}}{(2\pi)^3} \textrm{Tr} \ln {\cal G}_A^{-1}(i\omega_n,{\bf p})
\end{equation}
with the ${\bf A}$-dependent inverse propagator
\begin{equation}
{\cal G}_A^{-1}(i\omega_n,{\bf p})=\left(\begin{array}{cc}
i\omega_n-\epsilon_{{\bf p}+eQ_a{\bf A}}^a&\Delta
\\ \Delta&i\omega_n+\epsilon_{{\bf p}-eQ_b{\bf A}}^b\end{array}\right).
\end{equation}
Using the same trick in Section \ref{s3}, we have the relation
\begin{equation}
{\cal G}_A^{-1}={\cal G}^{-1}-e\Sigma_p{\bf p}\cdot{\bf
A}-{e^2\over 2}\Sigma_d{\bf A}^2
\end{equation}
and the derivative expansion
\begin{eqnarray}
&&\textrm{Tr}\ln {\cal G}_A^{-1}-\textrm{Tr}\ln {\cal G}^{-1}
=e{\bf
p}\cdot{\bf A} \textrm{Tr}\left({\cal G}\Sigma_p\right)\\
&&-{e^2\over 2}{\bf A}^2\textrm{Tr}\left({\cal G}\Sigma_d\right)
-\frac{e^2}{2}({\bf p}\cdot{\bf A})^2 \textrm{Tr}\left({\cal
G}\Sigma_p{\cal G}\Sigma_p\right)+\cdots. \nonumber
\end{eqnarray}
The Meissner mass squared can be read from the quadratic terms in
${\bf A}$. After some algebras we obtain
\begin{eqnarray}
m_A^2&&=e^2\left(\frac{n_a}{m_a}Q_a^2+\frac{n_b}{m_b}Q_b^2\right)\\
&&+\ e^2\int\frac{d^3{\bf
p}}{(2\pi)^3}\frac{p^2}{3}\left(\frac{Q_a^2}{m_a^2}{\cal
T}_{11}+\frac{Q_b^2}{m_b^2}{\cal
T}_{22}+\frac{2Q_aQ_b}{m_am_b}{\cal T}_{12}\right). \nonumber
\end{eqnarray}
The second term is just the long-wave and static limit of the
current-current correlation function. With the result of frequency
summations in Section \ref{s3}, we obtain an explicit expression,
\begin{widetext}
\begin{eqnarray}
m_A^2&=&e^2\left(\frac{n_a}{m_a}Q_a^2+\frac{n_b}{m_b}Q_b^2\right)+e^2\int{d^3{\bf
p}\over (2\pi)^3}{p^2\over 3}\times\nonumber\\
&&\Bigg[\left({Q_a\over m_a}-{Q_b\over
m_b}\right)^2u_p^2v_p^2{f(\epsilon_{\bf p}^A)+f(\epsilon_{\bf
p}^B)-1\over \epsilon_\Delta}+\left({Q_a\over m_a}u_p^2+{Q_b\over
m_b}v_p^2\right)^2f'(\epsilon_{\bf p}^A)+\left({Q_a\over
m_a}v_p^2+{Q_b\over m_b}u_p^2\right)^2f'(\epsilon_{\bf
p}^B)\Bigg]\ .
\end{eqnarray}
The formula is invariant under the exchange $a\leftrightarrow b$.
For $\Delta=0$, it is reduced to
\begin{equation}
m_A^2=e^2\int_0^\infty dp{p^2\over
2\pi^2}\left[\frac{f(\epsilon_p^a)}{m_a}Q_a^2+\frac{f(\epsilon_p^b)}{m_b}Q_b^2\right]+e^2\int_0^\infty
dp{p^4\over
6\pi^2}\Bigg[\frac{f'(\epsilon_p^a)}{m_a^2}Q_a^2+\frac{f'(\epsilon_p^b)}{m_b^2}Q_b^2\Bigg]
\end{equation}
\end{widetext}
and vanishes in the normal state, as we can expect.

The formulae of the superfluid density and the Meissner mass
squared we derived seem quite different. In the symmetric case
with $m_a=m_b\equiv m$ and $Q_a=Q_b=1$, we recover the well know
result \cite{fetter,nao}
\begin{eqnarray}
\rho_s&=&mn+\int\frac{d^3{\bf
p}}{(2\pi)^3}\frac{p^2}{3}\left[f^\prime(\epsilon_{\bf p}^A)+f^\prime(\epsilon_{\bf p}^B)\right], \nonumber\\
m_A^2&=&\frac{ne^2}{m}+\frac{e^2}{m^2}\int\frac{d^3{\bf
p}}{(2\pi)^3}\frac{p^2}{3}\left[f^\prime(\epsilon_{\bf
p}^A)+f^\prime(\epsilon_{\bf p}^B)\right],
\end{eqnarray}
and the proportional relation
\begin{eqnarray}
\frac{\rho_s}{m_A^2}=\frac{m^2}{e^2}
\end{eqnarray}
at any temperature $T<T_c$. In fact, one can easily check that for
systems with $Q_a/m_a=Q_b/m_b$, the proportional relation still
holds. However, for general asymmetric systems, this relation is
broken down.

\section {The Breached Pairing State}
\label{s5}
We calculate the superfluid density and the Meissner mass squared
for the breached pairing state at zero temperature in this
section. Explicitly, the dispersions of the fermionic
quasiparticles can be expressed as
\begin{equation}
\epsilon_{\vec p}^{A,B} = \sqrt{\left({p^2\over
2m}-\mu\right)^2+\Delta^2}\pm\left({p^2\over
2m^\prime}+\delta\mu\right)
\end{equation}
with the reduced masses $m=2m_am_b/(m_a+m_b)$ and
$m^\prime=2m_am_b/(m_b-m_a)$ and chemical potentials
$\mu=(\mu_a+\mu_b)/2$ and $\delta\mu=(\mu_b-\mu_a)/2$. One can
easily check that for
\begin{equation}
\Delta<\Delta_c=\frac{|p_b^2-p_a^2|}{4\sqrt{m_am_b}} ,
\end{equation}
with $p_i=\sqrt{2m_i\mu_i}$, one branch of the fermionic
quasiparticles can cross the momentum axis and hence becomes
gapless. This is the so called breached pairing state or interior
gap state\cite{liu}. The gapless nodes determined by
$\epsilon_{\bf p}^A=0$ or $\epsilon_{\bf p}^B=0$ are located at
$p=p_{1,2}$ with
\begin{equation}
p_{1,2}^2=\frac{p_a^2+p_b^2}{2}\mp\frac{1}{2}\sqrt{(p_a^2-p_b^2)^2-16m_am_b\Delta^2}.
\end{equation}
The gap equation which determines $\Delta$ at zero temperature
reads
\begin{equation}
-\frac{m}{4\pi a_s}=\int\frac{d^3{\bf
p}}{(2\pi)^3}\left[\frac{\Theta(\epsilon_{\bf
p}^A)-\Theta(-\epsilon_{\bf
p}^B)}{2\epsilon_\Delta}-\frac{1}{2\xi_p}\right]=0.
\end{equation}
Here $\Theta(x)$ is the step function, and the s-wave scattering
length $a_s$ is related to the bare coupling $g$ via
\begin{equation}
\frac{m}{4\pi a_s}=-\frac{1}{g}+\int\frac{d^3{\bf
p}}{(2\pi)^3}\frac{1}{2\xi_p}
\end{equation}
with $\xi_p=p^2/2m$. One can employ another regularization scheme
where the pairing interaction is restricted in a narrow region
around the common Fermi surface $p_F=\sqrt{2m\mu}$\cite{liu} which
is suitable only for weak coupling case. We have checked that the
results from these two regularization schemes are the same in weak
coupling.

In the BCS phase, all fermionic excitations are gapped, the gap
equation can be reduced to
\begin{equation}
-\frac{m}{4\pi a_s}=\int_0^\infty
dp\frac{p^2}{2\pi^2}\left(\frac{1}{2\epsilon_\Delta}-\frac{1}{2\xi_p}\right),
\end{equation}
and the solution in weak coupling is
\begin{equation}
\Delta_0\simeq 8e^{-2}\mu e^{-\frac{\pi}{2p_F|a_s|}}.
\end{equation}
Since all fermionic quasiparticles are gapped, we have
$n_a=n_b=n/2$, and the superfluid density reads
\begin{eqnarray}
\rho_s=m_an_a+m_bn_b=\frac{1}{2}(m_a+m_b)n,
\end{eqnarray}
which means that all fermions participate in the superfluid. In
weak coupling, we have approximately $n\simeq
\frac{p_F^3}{3\pi^2}$ and
\begin{eqnarray}
\int{d^3{\bf p}\over (2\pi)^3}{p^2\over 12}{\Delta^2\over
\epsilon_\Delta^3}\simeq\frac{mp_F^3}{12\pi^2}
\end{eqnarray}
which leads to
\begin{eqnarray}
m_A^2=\frac{ne^2(Q_a+Q_b)^2}{2(m_a+m_b)}.
\end{eqnarray}
Note that $m_A^2$ depends only on $Q_a+Q_b$, and this conclusion
is valid only in weak coupling we considered by hand.

In the breached pairing state with gapless fermionic excitations,
the gap equation can be expressed as
\begin{equation}
-\frac{m}{4\pi a_s}=\int_0^\infty
dp\frac{p^2}{2\pi^2}\left(\frac{1}{2\epsilon_\Delta}-\frac{1}{2\xi_p}\right)-\int_{p_1}^{p_2}dp\frac{p^2}{2\pi^2}\frac{1}{2\epsilon_\Delta}.
\end{equation}
Using the equation for the BCS gap $\Delta_0$, the solution of the
gap equation can be well described by\cite{caldas}
\begin{equation}
\Delta=\sqrt{\Delta_0(2\Delta_c-\Delta_0)}.
\end{equation}
The gap varies in the region $0<\Delta<\Delta_0$, and the
difference between the Fermi momenta satisfies
\begin{equation}
2\sqrt{m_am_b}\Delta_0<|p_a^2-p_b^2|<4\sqrt{m_am_b}\Delta_0.
\end{equation}
If we define the density asymmetry $\alpha=|n_a-n_b|/(n_a+n_b)$,
then $\Delta=\Delta_0$ corresponds to $\alpha=0$ and $\Delta=0$
corresponds to the maximal asymmetry $\alpha_c$. When $\alpha$
varies from $0$ to $\alpha_c$, the gap $\Delta$ decreases from
$\Delta_0$ to $0$.

We firstly discuss the case $m_a=m_b$ where the superfluid density
is proportional to the Meissner mass squared. In this case we have
$\Delta_c=\delta\mu$ and $p^2/2m^\prime=0$. Let's set
$\delta\mu>0$ without loss of generality. At zero temperature the
superfluid density reads
\begin{equation}
\label{kq3} \rho_s=mn-\frac{1}{6\pi^2}\int_0^\infty
dpp^4\delta(\epsilon_\Delta-\delta\mu).
\end{equation}
In the weak coupling region, $\epsilon_\Delta-\delta\mu=0$ has two
possible roots $p_1,p_2$ and $\rho_s$ can be evaluated as
\begin{equation}
\label{kq4}
\rho_s=mn\left[1-\eta\frac{\delta\mu\Theta(\delta\mu-\Delta)}{\sqrt{\delta\mu^2-\Delta^2}}\right]
\end{equation}
with the coefficient
$\eta=\left(p_1^3+p_2^3\right)/\left(6\pi^2n\right)$. Since the
coefficient $\eta$ is approximately equal to $1$, $\rho_s$ can be
well approximated by
\begin{equation}
\label{kq5} \rho_s \simeq
mn\left[1-\frac{\delta\mu\Theta(\delta\mu-\Delta)}{\sqrt{\delta\mu^2-\Delta^2}}\right].
\end{equation}
It is now clear that in the BP state with $\Delta<\delta\mu$,
$\rho_s$ becomes negative and the BP state is unstable. We should
emphasis that the function in the bracket is universal for gapless
fermion superfluids in equal mass systems. This function appears
in the Meissner mass squared for the 8th gluon in two flavor
gapless color superconductor\cite{huang3,huang4}. In some
anisotropic states in equal mass systems such as the LOFF
state\cite{giannakis2,he2} and the BP state via p-wave
pairing\cite{pwave}, a similar function appears. In the LOFF state
$\delta\mu$ is replaced by an angle dependent mismatch
$\delta_\theta$\cite{giannakis2,he2}, and in the p-wave BP state
the gap $\Delta$ is replaced by an anisotropic gap function
$\Delta_{\bf n}$\cite{pwave}.

In general case with $m_a\neq m_b$, we define a mass ratio
$\lambda=m_b/m_a$ and set $\lambda>1$ without loss of generality.
For $\lambda\neq1$, the results for $p_a<p_b$ and $p_a>p_b$(or
$n_a<n_b$ and $n_a>n_b$) are not symmetric. We will discuss these
two cases separately at zero temperature. For the sake of
simplicity, we set $Q_a=Q_b=1$ in the calculations.

\subsection {$p_a<p_b$($n_a<n_b$)}
In this case, the branch $\epsilon_{\bf p}^B$ becomes gapless and
we have $n_a({\bf p})=0,n_b({\bf p})=1$ in the region $p_1<p<p_2$.
At zero temperature, the superfluid density in the BP state can be
evaluated as
\begin{eqnarray}
\label{rhos}\rho_s=m_a\frac{\alpha_s p_1^3+\beta_s p_2^3}{6\pi^2}
\end{eqnarray}
with the coefficients $\alpha_s$ and $\beta_s$ defined as
\begin{eqnarray}
\alpha_s&=&1-\frac{\lambda}{|1-(\lambda+1)v_1^2|}-3(\lambda+1)\int_{R_1}dp\frac{p^2}{p_1^3}u_p^2,\nonumber\\
\beta_s&=&\lambda-\frac{\lambda}{|1-(\lambda+1)v_2^2|}+3(\lambda+1)\int_{R_2}dp
\frac{p^2}{p_2^3}v_p^2,
\end{eqnarray}
where $v_1^2$ and $v_2^2$ are the values of $v_p^2$ at $p=p_1$ and
$p=p_2$, and the integral regions $R_1$ and $R_2$ are $0<p<p_1$
and $p_2<p<\infty$ respectively. The Meissner mass squared in the
BP state can be evaluated as
\begin{eqnarray}
\label{mA}m_A^2=\frac{e^2}{m_b}\frac{\alpha_m p_1^3+\beta_m
p_2^3-\gamma_m p_F^3}{6\pi^2}
\end{eqnarray}
where the coefficients $\alpha_m,\beta_m$ and $\gamma_m$ are
defined as
\begin{eqnarray}
\alpha_m&=&\lambda-\frac{\left[1+(\lambda-1)v_1^2\right]^2}{|1-(\lambda+1)v_1^2|}-3(\lambda+1)\int_{R_1}dp\frac{p^2}{p_1^3}u_p^2,\nonumber\\
\beta_m&=&1-\frac{\left[1+(\lambda-1)v_2^2\right]^2}{|1-(\lambda+1)v_2^2|}+3(\lambda+1)\int_{R_2}dp\frac{p^2}{p_2^3}v_p^2,\nonumber\\
\gamma_m&=&\frac{(\lambda-1)^2}{\lambda+1}\int_{R_1+R_2}dp\frac{p^2}{p_F^3}
\frac{\Delta^2\xi_p}{[(\xi_p-\mu)^2+\Delta^2]^{3/2}} .
\end{eqnarray}

\begin{figure}
\centering
\includegraphics[width=7cm]{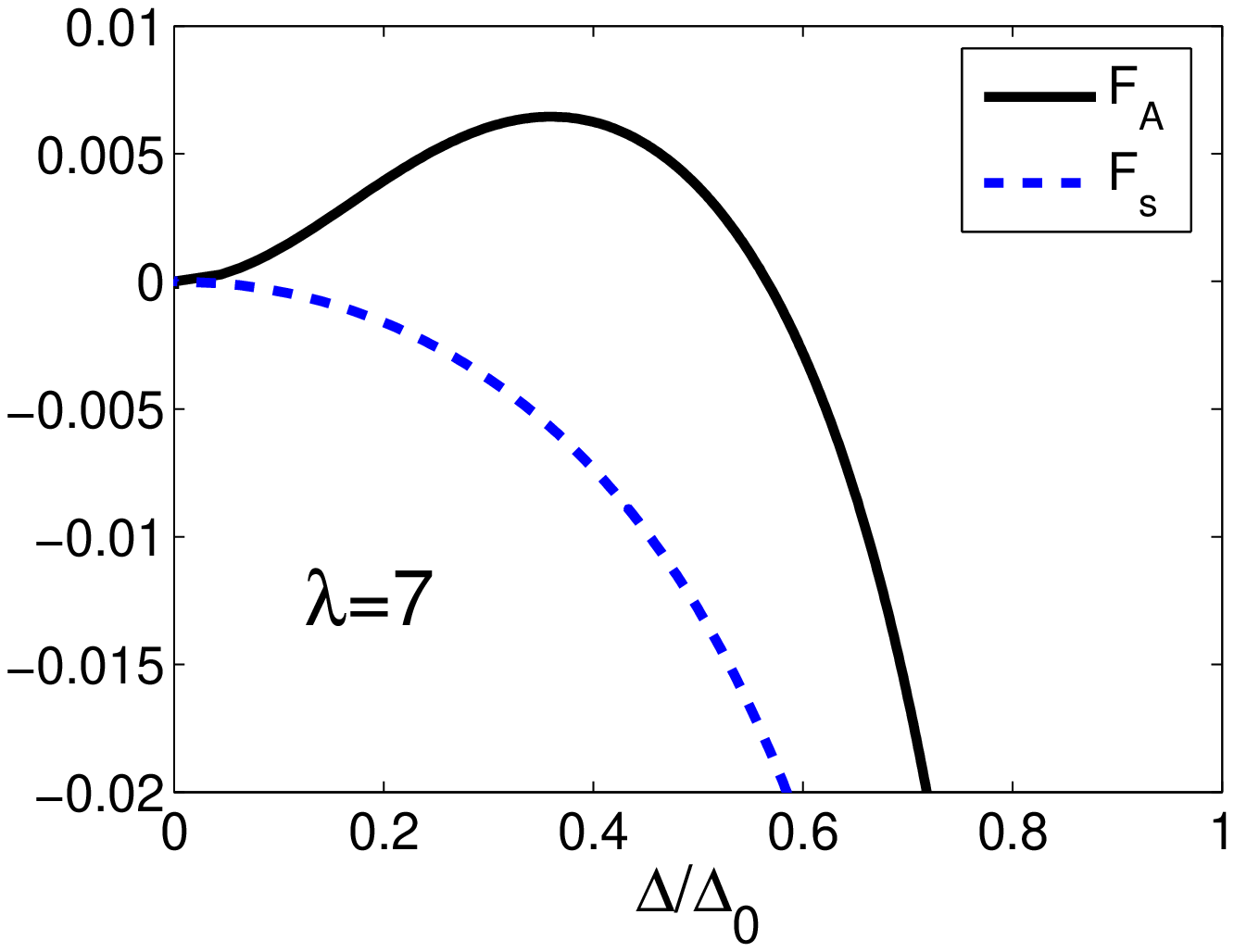}
\includegraphics[width=7cm]{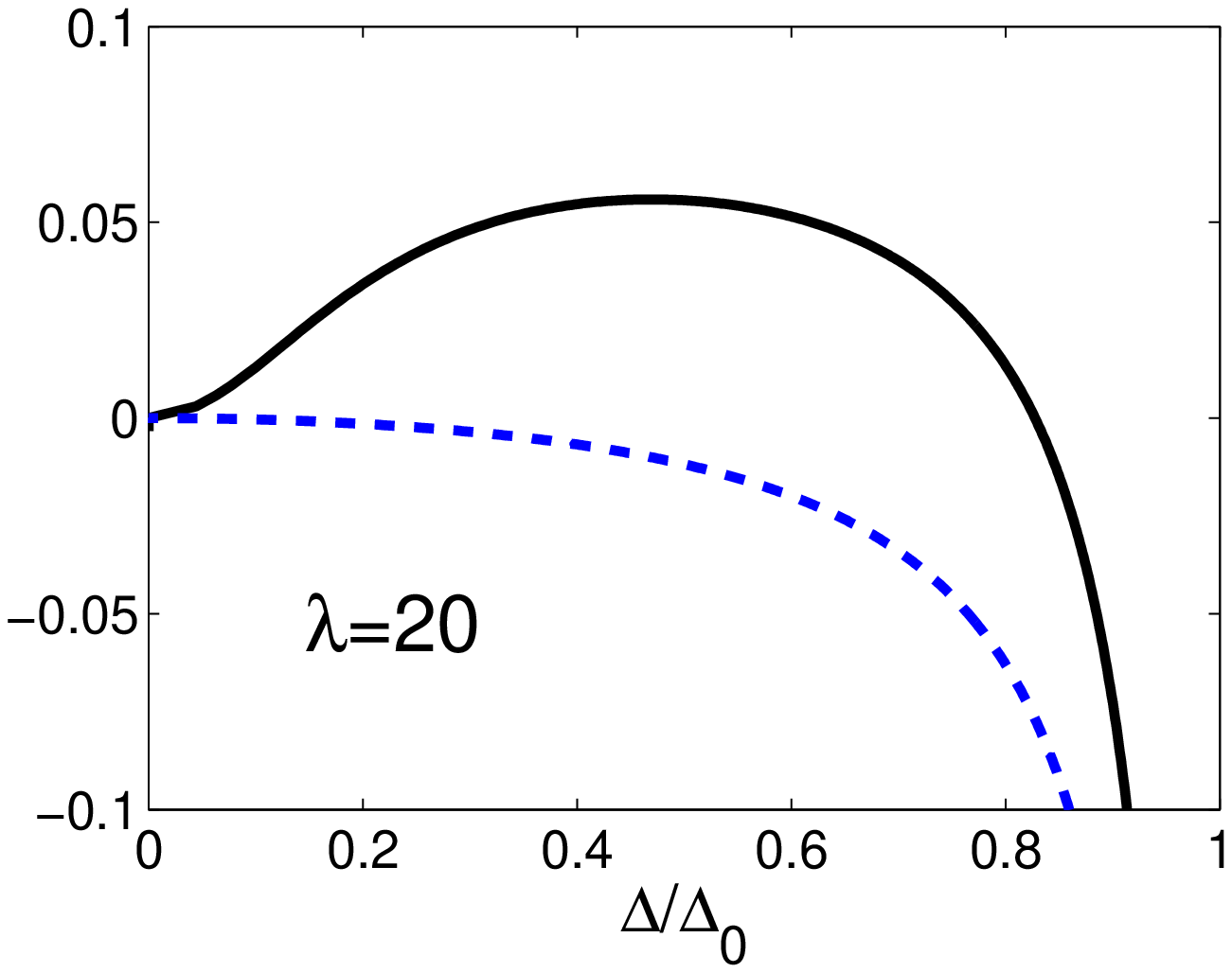}
\includegraphics[width=7cm]{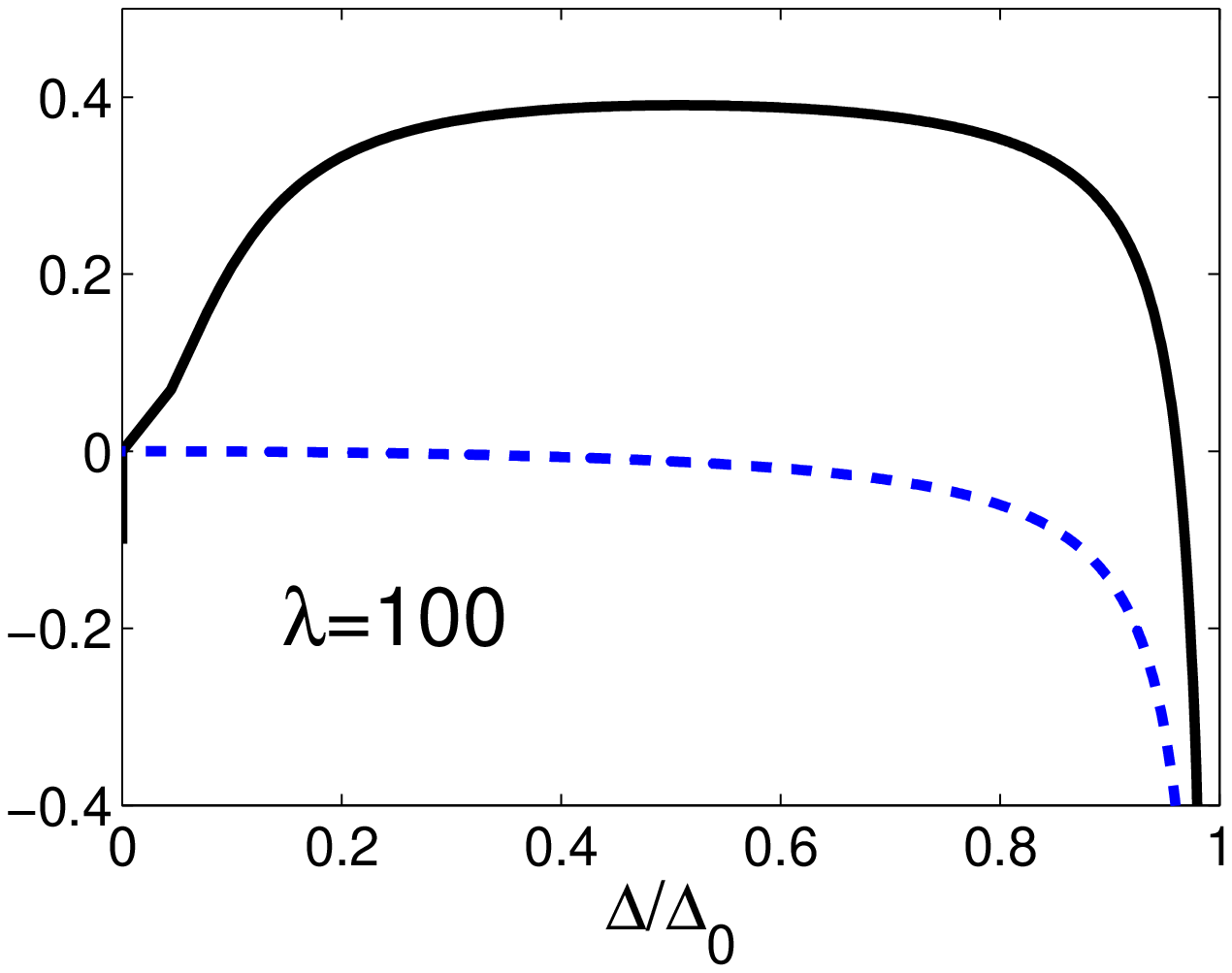}
\caption{The scaled Meissner mass squared $F_A=m_A^2/m_0^2$(solid
lines) and superfluid density $F_s=\rho_s/\rho_0$(dashed lines) as
functions of $\Delta/\Delta_0$ for different values of mass ratio
$\lambda$ in the case $p_a<p_b$. The normalization constants
$m_0^2$ and $\rho_0$ are chosen to be
$m_0^2=e^2m^2p_b^3/(m_a^2m_b)$ and $\rho_0=m_bp_b^3$. The BCS gap
$\Delta_0$ is chosen to be $\Delta_0=0.01\mu_b$. \label{fig1}}
\end{figure}

From $p_a<p_b$, we have $\lambda\mu_b>\mu_a$, and the chemical
potentials in the BP state satisfy
\begin{equation}
\frac{\Delta_0}{2}<\frac{\lambda\mu_b-\mu_a}{2\sqrt{\lambda}}<\Delta_0.
\end{equation}
Without loss of generality, we can keep $\mu_b$ fixed. After a
simple algebra we find that
\begin{equation}
\lambda\mu_b-2\sqrt{\lambda}\Delta_0<\mu_a<\lambda\mu_b-\sqrt{\lambda}\Delta_0.
\end{equation}
The lower bound corresponds to $\Delta=\Delta_0$ where $\alpha=0$,
and the upper bound corresponds to $\Delta=0$ where
$\alpha=\alpha_c$. Then we can calculate the superfluid density
and the Meissner mass squared as functions of $\Delta/\Delta_0$ in
the BP range $0<\Delta/\Delta_0<1$. In Fig.\ref{fig1}, we show the
superfluid density and Meissner mass squared for different values
of mass ratio $\lambda$. We found that the superfluid density is
always negative at any mass ratio, but the Meissner mass squared
is positive in the region $0<\Delta/\Delta_0<\nu$ with $\nu<1$.
When the mass ratio becomes very large, such as $\lambda=100$,
$\nu$ is close to $1$. Even though there exists a big room where
the Meissner mass squared is positive, both the superfluid density
and the Meissner mass squared tend to negative infinity near
$\Delta/\Delta_0=1$. Such a divergence at the BP-BCS transition
point, which comes from the divergent density of state of the
gapless excitations, can not be avoided\cite{wu,he}.

\begin{figure}
\centering
\includegraphics[width=7cm]{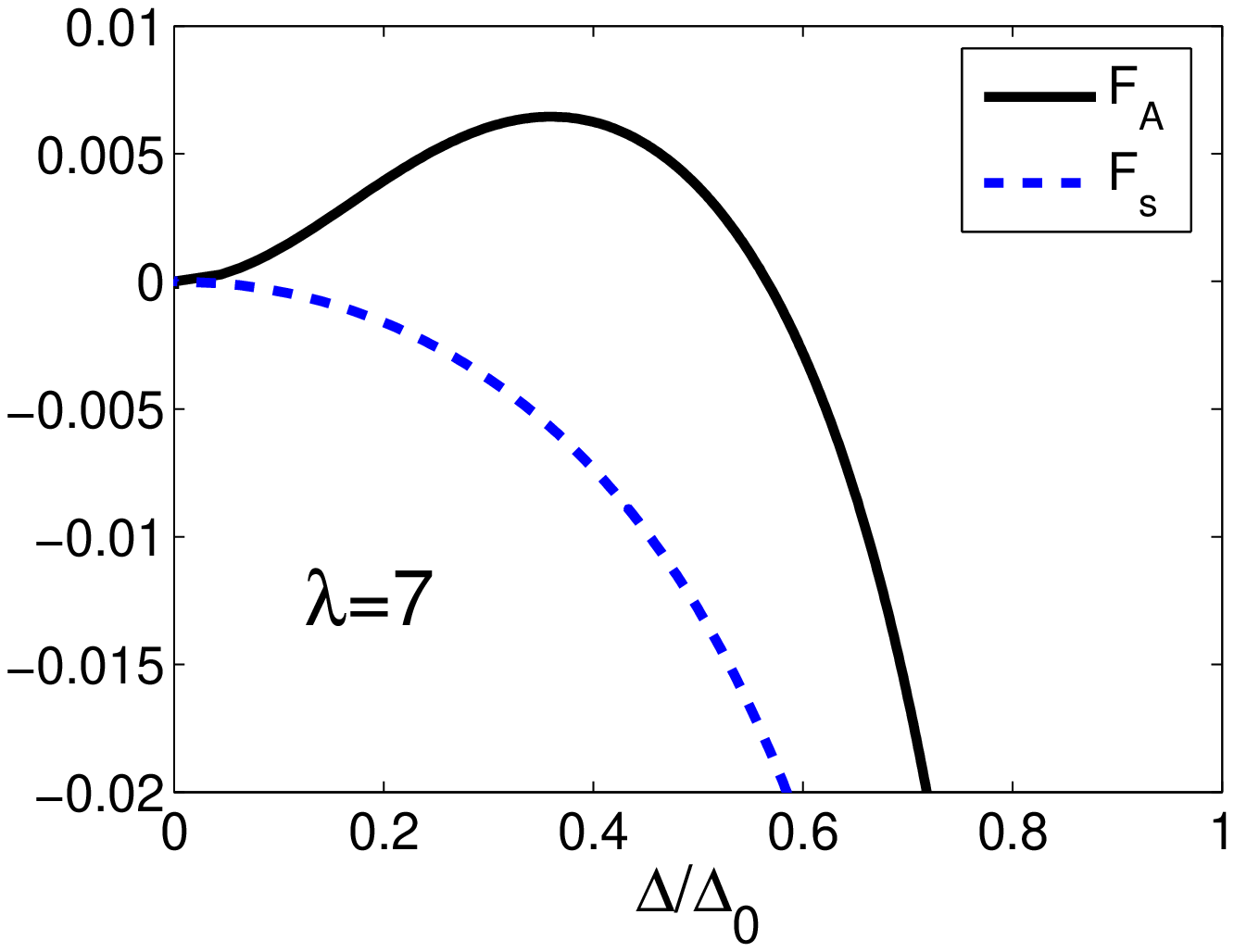}
\includegraphics[width=7cm]{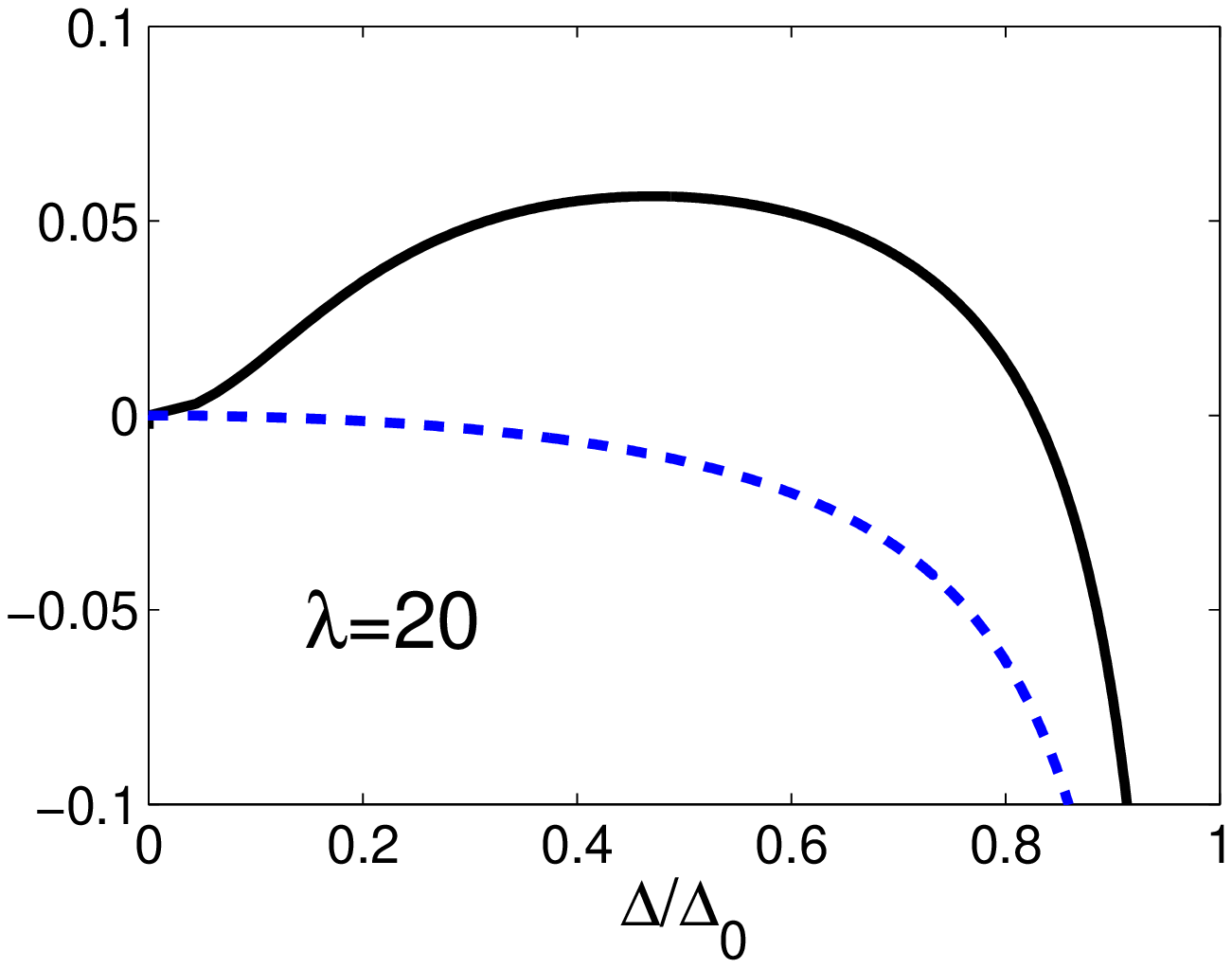}
\includegraphics[width=7cm]{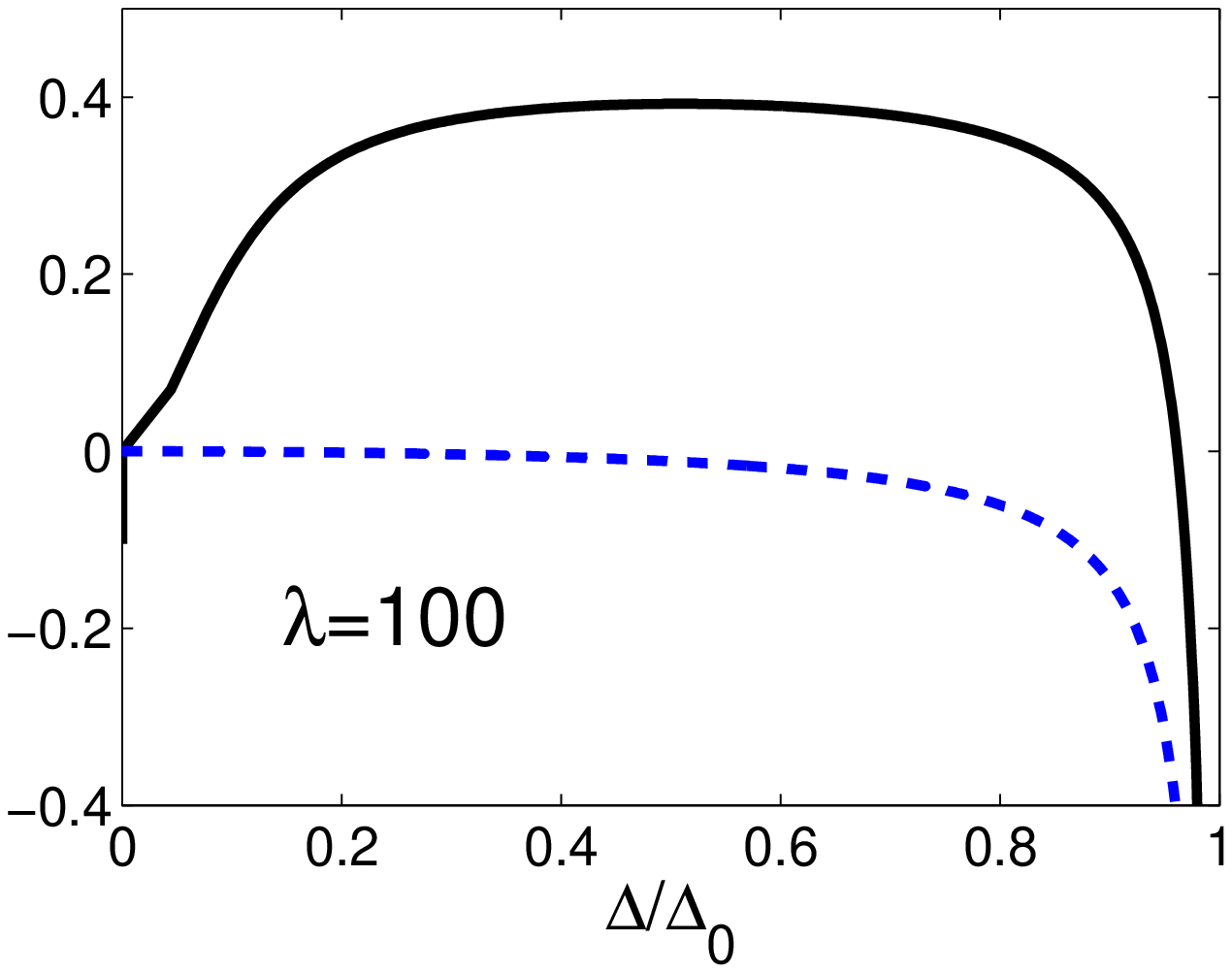}
\caption{The scaled Meissner mass squared $F_A$(solid lines) and
superfluid density $F_s$(dashed lines) as functions of
$\Delta/\Delta_0$ for different values of mass ratio $\lambda$ in
the case $p_a>p_b$. The BCS gap $\Delta_0$ is chosen to be
$\Delta_0=0.01\mu_b$. \label{fig2}}
\end{figure}

\subsection {$p_a>p_b$($n_a>n_b$)}
In this case, the branch $\epsilon^A_{\bf p}$ becomes gapless and
we have $n_a({\bf p})=1,n_b({\bf p})=0$ in the region $p_1<p<p_2$.
At zero temperature, the superfluid density in the BP state takes
the same form (\ref{rhos}) but with different coefficients
\begin{eqnarray}
\alpha_s&=&\lambda-\frac{\lambda}{|\lambda-(\lambda+1)v_1^2|}-3(\lambda+1)\int_{R_1}dp\frac{p^2}{p_1^3}u_p^2,\nonumber\\
\beta_s&=&1-\frac{\lambda}{|\lambda-(\lambda+1)v_2^2|}+3(\lambda+1)\int_{R_2}
dp\frac{p^2}{p_2^3}v_p^2.
\end{eqnarray}
The Meissner mass squared takes also the same form (\ref{mA}), but
the coefficients $\alpha_m,\beta_m$ are modified to
\begin{eqnarray}
\alpha_m&=&1-\frac{\left[\lambda-(\lambda-1)v_1^2\right]^2}{|\lambda-(\lambda+1)v_1^2|}-3(\lambda+1)\int_{R_1}dp\frac{p^2}{p_1^3}u_p^2,\nonumber\\
\beta_m&=&\lambda-\frac{\left[\lambda-(\lambda-1)v_2^2\right]^2}{|\lambda-(\lambda+1)v_2^2|}+3(\lambda+1)\int_{R_2}dp
\frac{p^2}{p_2^3}v_p^2,\nonumber\\
\end{eqnarray}
and $\gamma_m$ remains unchanged.

Similarly, for fixed $\mu_b$, we have
\begin{equation}
\lambda\mu_b+\sqrt{\lambda}\Delta_0<\mu_a<\lambda\mu_b+2\sqrt{\lambda}\Delta_0
\end{equation}
for the BP state. The superfluid density and the Meissner mass
squared are calculated in Fig.\ref{fig2} as functions of
$\Delta/\Delta_0$ in the range $0<\Delta/\Delta_0<1$. The
qualitative behavior is almost the same as in the case $p_a<p_b$.

In summary, we have shown in weak coupling that, the superfluid
density of the BP state is always negative, while the Meissner
mass squared can be positive in a wide region. The conclusion here
is valid for stronger coupling, if there exist two gapless
nodes\cite{smit}.

\section {LOFF State}
\label{s6}
In weak coupling the superfluid density of BP state is always
negative, which indicates that the BP state is unstable and some
inhomogeneous and anisotropic state is energetically favored. In
this section we show that the LOFF state is energetically favored
due to the negative superfluid density.

For the sake of simplicity, we consider the simplest pattern of
LOFF state, namely the single plane wave ansatz
\begin{equation}
\langle\phi(x)\rangle=\Delta e^{2i{\bf q}\cdot{\bf x}}\ ,\ \ \
\langle\phi^*(x)\rangle=\Delta e^{-2i{\bf q}\cdot{\bf x}},
\end{equation}
where $\Delta$ is a real quantity, and $2{\bf q}$ is the so called
LOFF momentum which is the total momentum of a Cooper pair. To
evaluate the thermodynamic potential of the LOFF state, we often
define new fermion fields $\chi_a(x)=e^{i{\bf q}\cdot{\bf
x}}\psi_a(x)$ and $\chi_b(x)=e^{i{\bf q}\cdot{\bf x}}\psi_b(x)$,
and then can directly evaluate the Gaussian path integral in the
new basis $\chi_a,\chi_b$. Following this way, the thermodynamic
potential reads
\begin{equation}
\Omega =\frac{\Delta^2}{g}-T\sum_n\int\frac{d^3{\bf
p}}{(2\pi)^3}\textrm{Tr}\ln {\cal G}_q^{-1}(i\omega_n,{\bf p})
\end{equation}
in terms of the new inverse propagator
\begin{equation}
{\cal G}_q^{-1}(i\omega_n,{\bf p})=\left(\begin{array}{cc}
i\omega_n-\epsilon_{{\bf p}+{\bf q}}^a&\Delta
\\ \Delta&i\omega_n+\epsilon_{{\bf p}-{\bf q}}^b\end{array}\right).
\end{equation}
This is just the $({\bf q}+{\bf p},{\bf q}-{\bf p})$ picture of
the LOFF pairing, which means the fermions in the cooper pair move
together with a total momentum $2{\bf q}$. To see whether the LOFF
state is energetically favored, we take the small $q$ expansion
\begin{equation}
\Omega(q)-\Omega(0)=\frac{1}{2}\frac{\partial^2 \Omega}{\partial
q^2}\bigg|_{q=0}q^2+O(q^4),
\end{equation}
where we have chosen a suitable z-direction such that ${\bf
q}=(0,0,q)$. Notice that a linear term in $q$ vanishes
automatically. One can easily observe the following relation
between the momentum susceptibility and the Meissner mass squared:
\begin{equation}
m_A^2=e^2\frac{\partial^2 \Omega}{\partial q^2}\Big|_{q=0}.
\end{equation}
For large mass difference, since the Meissner mass squared is
positive in the small $\Delta$ region which is just the window for
LOFF state, as we have shown in the last section, we may conclude
that the LOFF state is not energetically favored. However, we
shall argue in the following that this is not the truth.

To give a correct argument we focus on the fact that the
superfluid density is always negative for any mass ratio. Notice
that we can do any transformation like
\begin{equation}
\chi_a(x)=e^{i{\bf q}_a\cdot{\bf x}}\psi_a(x),\ \
\chi_b(x)=e^{i{\bf q}_b\cdot{\bf x}}\psi_b(x)
\end{equation}
to evaluate the effective potential, since the phase factor in the
condensate can be eliminated by any ${\bf q}_a$ and ${\bf q}_b$
satisfying ${\bf q}_a+{\bf q}_b=2{\bf q}$. For such a general
transformation, the thermodynamic potential reads
\begin{equation}
\Omega =\frac{\Delta^2}{g}-T\sum_n\int\frac{d^3{\bf
p}}{(2\pi)^3}\textrm{Tr}\ln {\cal G}_{q_a,q_b}^{-1}(i\omega_n,{\bf
p})
\end{equation}
with
\begin{equation}
{\cal G}_{q_a,q_b}^{-1}(i\omega_n,{\bf p})=\left(\begin{array}{cc}
i\omega_n-\epsilon_{{\bf p}+{\bf q}_a}^a&\Delta
\\ \Delta&i\omega_n+\epsilon_{{\bf p}-{\bf q}_b}^b\end{array}\right).
\end{equation}
This arbitrariness of phase transformation is directly linked to
the fact that the symmetry group of the model Lagrangian is
$U(1)_{\varphi_a}\otimes U(1)_{\varphi_b}$. To link the LOFF state
and the superfluid density, we introduce a LOFF velocity ${\bf w}$
such that
\begin{equation}
{\bf q}_a=m_a{\bf w}, \ \ {\bf q}_b=m_b{\bf w}.
\end{equation}
With a suitable choice of coordinates such that ${\bf w}=(0,0,w)$,
we can do the similar small $w$ expansion
\begin{equation}
\Omega(w)-\Omega(0)=\frac{1}{2}\frac{\partial^2 \Omega}{\partial
w^2}\Big|_{w=0}w^2+O(w^4).
\end{equation}
Also, one can easily observe the following relation between the
velocity susceptibility and the superfluid density:
\begin{equation}
\rho_s=\frac{\partial^2 \Omega}{\partial w^2}\Big|_{w=0}.
\end{equation}
This intuitive argument indicates that the energetically favored
momentum configuration of LOFF state is ${\bf q}_a=m_a{\bf w},
{\bf q}_b=m_b{\bf w}$, which means the single plane wave LOFF
ansatz is nothing but the anisotropic state with spontaneously
generated superflow ${\bf v}_s$ if we identify ${\bf w}={\bf
v}_s$. In fact, it is quite easy for us to understand this fact.
The physical picture of the LOFF state is that the fermions in a
cooper pair move together with a nonzero momentum, and hence they
should possess a same velocity, not momentum. In fact, we have
checked numerically that for general choice of ${\bf q}_a$ and
${\bf q}_b$ that the quadratic term in the expansion is always
negative only when ${\bf q}_a=m_a{\bf w}, {\bf q}_b=m_b{\bf w}$.

With the proper configuration ${\bf q}_a=m_a{\bf w}, {\bf
q}_b=m_b{\bf w}$, we can evaluate the effective potential as
\begin{eqnarray}
\Omega(\Delta,w)&=&\frac{\Delta^2}{g}-\int
\frac{d^3{\bf p}}{(2\pi)^3}(E_\Delta-E_S)\nonumber\\
&-&\int \frac{d^3{\bf p}}{(2\pi)^3}\left[H(E^A_{\bf p})+H(E^B_{\bf
p})\right].
\end{eqnarray}
Here $H(x)=T\ln(1+e^{-x/T})$, $E_S=\xi_p-\mu_w$,
$E_\Delta=\sqrt{E_S^2+\Delta^2}$ and $E_{\bf p}^A, E_{\bf p}^B$
are the energies of the quasiparticles
\begin{eqnarray}
E_{\bf
p}^{A,B}=E_\Delta\pm\left(\frac{p^2}{2m^\prime}+\delta_w+{\bf
p}\cdot{\bf w}\right)
\end{eqnarray}
with $\mu_w=\mu-(m_a+m_b)w^2/4$ and
$\delta_w=\delta\mu+(m_a-m_b)w^2/4$. Notice that a new kinetic
energy term $(m_a-m_b)w^2/4$ which vanishes in the equal mass case
appears in the asymmetric part of the quasiparticle dispersions.
In fact, this momentum configuration is the most convenient one to
calculate the LOFF solution, since the anisotropic term ${\bf
p}\cdot{\bf w}$ appears only in the asymmetric part.

Now we give a preliminary discussion on the LOFF state with mass
difference. For convenience, we assume the pairing interaction is
restricted in the region $p_F-\Lambda<|\vec{p}|<p_F+\Lambda$ with
$\Lambda\ll p_F$, here $\Lambda$ serves as a natural ultraviolet
cutoff in the theory. In weak coupling we can safely neglect the
terms of order $O(w^2)$ and do the following replacement
\begin{eqnarray}
{\bf p}\cdot{\bf w}\rightarrow p_Fw\cos\theta,\ \ \ \
\frac{p^2}{2m^\prime}\rightarrow \frac{p_F^2}{2m^\prime},
\end{eqnarray}
where $\theta$ is angle between ${\bf p}$ and ${\bf w}$. Up to
now, all things become the same as those in the equal mass
systems\cite{takada,giannakis2}, and the conclusions there can be
directly applied. If the chemical potentials for the two species
are fixed, the corresponding LOFF window is
\begin{eqnarray}
0.707\Delta_0<\left|\delta\mu+\frac{p_F^2}{2m^\prime}\right|<0.754\Delta_0,
\end{eqnarray}
where $\Delta_0$ is the BCS gap, and the LOFF velocity $w$ is
approximately given by\cite{takada,giannakis2}
\begin{eqnarray}
p_Fw\simeq1.2\left|\delta\mu+\frac{p_F^2}{2m^\prime}\right|.
\end{eqnarray}
Defining the mass asymmetry $\epsilon=(m_b-m_a)/(m_b+m_a)$ we have
\begin{eqnarray}
\delta\mu+\frac{p_F^2}{2m^\prime}=\frac{1}{2}\left[(1+\epsilon)\mu_b-(1-\epsilon)\mu_a\right]\equiv\delta(\epsilon),
\end{eqnarray}
and then can reexpress the LOFF window as the conventional form in
equal mass case
\begin{eqnarray}
0.707\Delta_0<|\delta(\epsilon)|<0.754\Delta_0
\end{eqnarray}
and $p_Fw\simeq 1.2\delta(\epsilon)$. Due to the relation
\begin{eqnarray}
\delta(\epsilon)=\frac{p_b^2-p_a^2}{2(m_a+m_b)},
\end{eqnarray}
the size of the LOFF momentum is
\begin{eqnarray}
|2{\bf q}|=(m_a+m_b)w\simeq 1.2(p_b-p_a),
\end{eqnarray}
which is just we expect. A LOFF state induced by a pure mass
difference is of great interest since in some physical systems the
chemical potentials are always equal due to chemical equilibrium.
Setting $\mu_a=\mu_b\equiv\mu$ we obtain the mass difference
window
\begin{eqnarray}
0.707\frac{\Delta_0}{\mu}<|\epsilon|<0.754\frac{\Delta_0}{\mu}.
\end{eqnarray}
In weak coupling, $\Delta_0\ll\mu$, the LOFF state can exist only
when the mass asymmetry is very small, which may be realized in
electronic systems. If the particle number densities $n_a$ and
$n_b$ are fixed, we should compare the free energy ${\cal
F}=\Omega+\mu_an_a+\mu_bn_b$. In this case we have the similar
expansion ${\cal F}(w)={\cal F}(0)+\rho_sw^2/2+O(w^4)$ which means
the LOFF state is more stable than BP state. The LOFF window will
be larger, which is similar to the equal mass system\cite{he2,hu}.
Such a situation may be realized in cold atomic Fermi gas, such as
a mixture of $^6$Li and $^{40}$K atoms.

Finally, we calculate the superfluid density tensor and the
Meissner mass squared tensor in the LOFF state.  Since the
rotational symmetry $O(3)$ is broken down to $O(2)$, the
superfluid density and Meissner mass squared become tensors
$\rho_{ij}$ and $(m_A^2)_{ij}$. We can decompose them into a
transverse part and a longitudinal part
\begin{eqnarray}
\rho_{ij}=\rho_T(\delta_{ij}-\hat{w}_i\hat{w}_j)+\rho_L\hat{w}_i\hat{w}_j,\nonumber\\
(m_A^2)_{ij}=m^2_T(\delta_{ij}-\hat{w}_i\hat{w}_j)+m^2_L\hat{w}_i\hat{w}_j
\end{eqnarray}
with $\hat{w}\equiv{\bf w}/|{\bf w}|$. The transverse and
longitudinal superfluid density read
\begin{eqnarray}
\rho_T&=&m_an_a+m_bn_b+\frac{3}{4}\int_{-1}^1d\cos\theta\sin^2\theta F(\cos\theta), \nonumber\\
\rho_L&=&m_an_a+m_bn_b-\frac{3}{2}\int_{-1}^1d\cos\theta\cos^2\theta
F(\cos\theta)\nonumber\\
\end{eqnarray}
with the function $F(\cos\theta)$ defined as
\begin{eqnarray}
F(\cos\theta)=\int_0^\infty
dp\frac{p^4}{4\pi^2}\left[f^\prime(E_{\bf p}^A)+f^\prime(E_{\bf
p}^B)\right],
\end{eqnarray}
while the transverse and longitudinal Meissner mass squared read
\begin{eqnarray}
m^2_T&=&e^2\left(\frac{n_a}{m_a}+\frac{n_b}{m_b}\right)+\frac{3e^2}{4}\int_{-1}^1d\cos\theta\sin^2\theta G(\cos\theta), \nonumber\\
m^2_L&=&e^2\left(\frac{n_a}{m_a}+\frac{n_b}{m_b}\right)+\frac{3e^2}{2}\int_{-1}^1d\cos\theta\cos^2\theta
G(\cos\theta)\nonumber\\
\end{eqnarray}
with the function $G(\cos\theta)$ defined as
\begin{eqnarray}
G(\cos\theta)&=&\int_0^\infty
dp\frac{p^4}{4\pi^2}\frac{4u_p^2v_p^2}{m^\prime}{f(E_{\bf
p}^A)+f(E_{\bf p}^B)-1\over
\epsilon_\Delta}\nonumber\\
&+&\int_0^\infty dp\frac{p^4}{4\pi^2}\left({u_p^2\over
m_a}+{v_p^2\over m_b}\right)^2f'(E_{\bf p}^A)\nonumber\\
&+&\int_0^\infty dp\frac{p^4}{4\pi^2}\left({v_p^2\over
m_a}+{u_p^2\over m_b}\right)^2f'(E_{\bf p}^B).
\end{eqnarray}
In equal mass systems, we have $m^2_T\propto \rho_T$ and can prove
tya5 they are both zero\cite{giannakis2}, which means that there
are no transverse Meissner effect and superfluid density. The
reason is that the formula of the transverse Meissner mass squared
is just the the gap equation for the LOFF momentum,
see\cite{takada,giannakis2}. However, for unequal mass systems,
the gap equation seems the same as equal mass systems, but the
formula of Meissner mass squared becomes quite different, and
there are both transverse and longitudinal Meissner effects.

\section {Extension to Finite Range Interaction} \label{s7}
The formulae for the superfluid density and Meissner mass squared
we derived are based on the point interaction model
(\ref{lagrangian}). In this section, we show that the formula can
be directly applied to finite range interaction systems, if we
replace the constant gap $\Delta$ by a momentum-dependent gap
function $\Delta({\bf p})$.

With a finite range interaction, the Lagrangian can be written as
\begin{eqnarray}
{\cal L}&=&\int d^3{\bf
x}\sum_{i=a,b}\psi_{i}^*({\bf x},\tau)\left(-\partial_\tau+\frac{\nabla^2}{2m_i}+\mu_i\right)\psi_{i}({\bf x},\tau)\nonumber\\
&+&\int d^3{\bf x} d^3{\bf y}\psi_{a}^*({\bf x})\psi_{b}^*({\bf
y})V({\bf x},{\bf y})\psi_{b}({\bf y})\psi_{a}({\bf x}),
\end{eqnarray}
where we have assumed that the interaction is static. For
convenience, we define the condensates
\begin{eqnarray}
\Phi({\bf x},{\bf y})&=&\langle\psi_{b}({\bf y})\psi_{a}({\bf x})\rangle,\nonumber\\
\Phi^*({\bf x},{\bf y})&=&\langle\psi_{a}^*({\bf
x})\psi_{b}^*({\bf y})\rangle,
\end{eqnarray}
and the gap functions
\begin{eqnarray}
\Delta({\bf x},{\bf y})&=&V({\bf x},{\bf y})\langle\psi_{b}({\bf y})\psi_{a}({\bf x})\rangle,\nonumber\\
\Delta^*({\bf x},{\bf y})&=&V({\bf x},{\bf
y})\langle\psi_{a}^*({\bf x})\psi_{b}^*({\bf y})\rangle.
\end{eqnarray}
If the system is translational invariant with $V({\bf x},{\bf
y})=V({\bf x}-{\bf y})$, $\Phi,\Delta$ and their complex
conjugates depend only on ${\bf x}-{\bf y}$. In mean field
approximation, the thermodynamic potential can be evaluated as
\begin{eqnarray}
\Omega &=&-T\sum_n\int\frac{d^3{\bf p}}{(2\pi)^3}\textrm{Tr} \ln
{\cal
G}^{-1}(i\omega_n,{\bf p})\nonumber\\
&+&\int\frac{d^3{\bf p}}{(2\pi)^3}\int\frac{d^3{\bf
q}}{(2\pi)^3}\Phi({\bf p})\Phi^*({\bf q})V({\bf p}-{\bf q})
\end{eqnarray}
in terms of the inverse fermion propagator
\begin{equation}
{\cal G}^{-1}(i\omega_n,{\bf p})=\left(\begin{array}{cc}
i\omega_n-\epsilon_{\bf p}^a&\Delta({\bf p})
\\ \Delta^*({\bf p})&i\omega_n+\epsilon_{\bf
p}^b\end{array}\right),
\end{equation}
where $V({\bf p}),\Phi({\bf p})$ and $\Delta({\bf p})$ are Fourier
transformation of $V({\bf x}-{\bf y}), \Phi({\bf x}-{\bf y})$ and
$\Delta({\bf x}-{\bf y})$. Since the derivation of the superfluid
density and Meissner mass squared depend only on the fermion
propagator ${\cal G}$, we conclude that the formulae for
superfluid density and Meissner mass squared derived in Sections
\ref{s5} and \ref{s6} are still valid in the finite range
interaction model, if we replace the constant gap $\Delta$ by the
momentum-dependent gap function $\Delta({\bf p})$.

The BP state with zero range interaction suffers negative
superfluid density, and is hence ruled out. It was proposed that
the BP state may be stable in a finite range interaction model
with large mass ratio\cite{forbes}, since it is the global minimum
of the thermodynamic potential with fixed chemical potentials. For
a complete study, we check now the superfluid density.

For a spherically symmetric potential $V(r)$, the gap function
depends only on $|{\bf p}|$ and satisfies the integral equation
\begin{equation}
\Delta(q)=\int\frac{d^3{\bf p}}{(2\pi)^3}V\left(|{\bf q}-{\bf
p}|\right)\frac{\Theta(\epsilon_{\bf p}^A)-\Theta(-\epsilon_{\bf
p}^B)}{2\sqrt{\left(\xi_p-\mu\right)^2+\Delta^2(p)}}\Delta(p).
\end{equation}
For a given potential $V$ and Fermi surface mismatch, we can solve
the equation and then determine the ground state with the lowest
thermodynamic potential. Once the BP solution is obtained, we can
calculate the superfluid density and Meissner mass squared.

For simplicity, let us concentrate on the superfluid density in
the case with $p_b>p_a$. The superfluid density is still in the
form (\ref{rhos}) but the coefficients become
\begin{eqnarray}
\alpha_s=1-\frac{\lambda}{|g(p_1)-(\lambda+1)v_1^2|}-3(\lambda+1)\int_0^{p_1}dp\frac{p^2}{p_1^3}u_p^2,\nonumber\\
\beta_s=\lambda-\frac{\lambda}{|g(p_2)-(\lambda+1)v_2^2|}+3(\lambda+1)\int_{p_2}^\infty
dp \frac{p^2}{p_2^3}v_p^2,\nonumber\\
\end{eqnarray}
where the function $g(p)$ is defined as
\begin{eqnarray}
g(p)=1+\frac{m_b}{p}\frac{\Delta(p)\Delta^\prime(p)}{\sqrt{\left(\xi_p-\mu\right)^2+\Delta^2(p)}},
\end{eqnarray}
with $\Delta^\prime(p)=d\Delta(p)/d p$. For the point interaction
model, we have $\Delta^\prime(p)=0$ and hence $g(p)=1$ which leads
to negative superfluid density, as we discussed above.

Generally, the gap function peaks at $p=p_0$ and drops down fast
for $p>p_0$. In this case, we have $g(p)=1$ and $v_p^2=0$ for
$p\geq p_2$\cite{forbes}, and the sign of the superfluid density
depends only on $\alpha_s$. The condition to produce a BP state
with positive superfluid density is then
\begin{equation}
|g(p_1)-(\lambda+1)v_1^2|>\frac{\lambda}{1-3(\lambda+1)p_1^{-3}\int_0^{p_1}dpp^2u_p^2}.
\end{equation}
If the slope of the gap function at $p=p_1$ is very large, the
condition can be easily satisfied.

For the interaction with a momentum cutoff
$p_\Lambda$\cite{forbes}, the momentum structure of the gap
function is $\Delta(p)=\Delta$ for $p<p_\Lambda$ and $\Delta(p)=0$
for $p>p_\Lambda$, and we have $g(p)= 1$ for all $p$ except at
$p=p_\Lambda$. Since in general case the positions of the zero
nodes are not exactly located at the cutoff, $p_{1,2}\neq
p_\Lambda$, the situation in this model is just the same as in the
point interaction model.

\section {Concluding Remarks}
\label{s8}
We have derived the superfluid density and the Meissner mass
squared for the fermion cooper pairing with unequal masses via a
standard field theory approach. For equal mass systems, the two
variables are indeed proportional to each other, while for unequal
mass systems, this relation breaks down. In the breached pairing
states with zero range interaction, the superfluid density is
always negative, but the Meissner mass squared is positive in a
wide region. As a consequence, the momentum configuration of the
LOFF pairing should be correctly established. We propose a proper
momentum configuration for LOFF pairing with unequal masses and
show that the single plane wave LOFF configuration in unequal mass
system is physically equivalent to an anisotropic state with
spontaneously generated superflow. These conclusions are valid
only in weak coupling. Whether they are valid for stronger
coupling, especially in the BCS-BEC crossover, should be examined.

There are some problems related to the arbitrariness in the phase
transformation induced by the $U(1)_{\varphi_a}\otimes
U(1)_{\varphi_b}$ symmetry. To investigate the Goldstone mode or
the phase fluctuation in the superfluid state, we often neglect
the fluctuation of the amplitude of the order parameter and write
$\phi(x)=\Delta e^{2i\theta(x)}$. Using the standard phase
transformation $\psi_{i}(x)=\tilde{\psi}_{i}(x)e^{i\theta(x)}$ we
can obtain the effective action for the Goldstone boson. However,
generally we can transform the fermion fields as
$\psi_{i}(x)=\tilde{\psi}_{i}(x)e^{i\nu_i\theta(x)}$ with $\nu_a$
and $\nu_b$ arbitrary constants satisfying the constraint
$\nu_a+\nu_b=2$. The low energy effective action for the phase
field $\theta(x)$ generally reads
\begin{equation}
S_{eff}[\theta]=-\frac{1}{2}\sum_q\left({\cal D}q_0^2-{\cal P}{\bf
q}^2\right)|\theta(q_0,{\bf q})|^2.
\end{equation}
Only when all fermionic excitations are gapped and at weak
coupling limit, we find ${\cal D}$ and ${\cal P}$ are independent
of $\nu_a$ and $\nu_b$. Hence the result of Goldstone boson
velocity in \cite{he3} is safe. This problem may imply that we can
not safely neglect the fluctuation of the amplitude of the order
parameter in strong coupling or in the gapless phases.

Another problem is the stability condition related to the phase
fluctuation. The superfluid density $\rho_s$ is often regarded as
a quantity to judge the stability of BP state\cite{pao}. For
simplicity we focus on the equal mass case. When $\rho_s$ is
negative, it directly means that the LOFF state has lower energy
than the BP state. However, this is true only for the standard
LOFF state with ${\bf q}_a={\bf q}_b$. For general case with ${\bf
q}_a=\nu_a{\bf q}$ and ${\bf q}_b=\nu_b{\bf q}$, we should check
the sign of $\kappa_q=\partial^2 \Omega/\partial q^2|_{q=0}$ for
all possible $\nu_a$ and $\nu_b$. While $\rho_s$ is positive in
strong coupling BEC region which means BP state is stable against
the standard LOFF state with ${\bf q}_a={\bf q}_b$\cite{pao},
there is no direct observation that $\kappa_q$ is positive for any
$\nu_a$ and $\nu_b$, such as $\nu_a=2,\nu_b=0$. If $\kappa_q$
becomes negative for $\nu_a\neq\nu_b$, a non-standard LOFF state
with ${\bf q}_a\neq{\bf q}_b$ is energetically favored in strong
coupling.

{\bf Acknowledgement:} We thank H.Caldas, M.M.Forbes, M.Huang,
W.V.Liu, H.Ren and H.Zhai for helpful discussions. The work was
supported in part by the grants NSFC10425810, 10435080, 10575058
and SRFDP20040003103.

\end{document}